# Decoherent many-body dynamics of a nano- mechanical resonator coupled to charge qubits

M. Abdel-Aty[1], J. Larson[2] and H. Eleuch[3]

[1] *University of Bahrain, 32038 Kingdom of Bahrain and Sohag University, Egypt*

[2] *NORDITA, Se-106 91 Stockholm, Sweden*

[3] *Institute for Quantum Studies, Texas A&M University, College Station Texas, U.S.A.*

## Abstract

The dynamics of charge qubits coupled to a nanomechanical resonator under influence of both a phonon bath in contact with the resonator and irreversible decay of the qubits is considered. The focus of our analysis is devoted to multi partite entanglement and the effects arising from the coupling to the reservoir. It is shown that despite losses, entanglement formation may still persist for relatively long times and it is especially robust against temperature dependence of the reservoir. Together with control of system parameters, the system may therefore be especially suited for quantum information processing. Furthermore, our results shed light on the evolution of open quantum many-body systems. For instance, due to intrinsic qubit-qubit couplings our model is related to a driven $XY$ spin model.

## I. INTRODUCTION

The interface between quantum and classical mechanics has been an issue of intense research ever since the very early days of quantum physics. Until recently, the topic was restricted to the theoretical community, but the experimental progress seen during the last decade has made it possible to now study quantum properties of objects with a large number of degrees of freedom. For instance, the double slit experiment with $C-60$ molecules has been realized[1], and the decoherence of superposition electromagnetic coherent states with 10 photons has been measured[2] (which later was improved to a situation with 36 photons[3]). A system that allows for experimental studies of the classical limit is an optomechanical cavity. Here a cavity mirror or a membrane interacts via radiation pressure with the cavity field. At the moment, utilizing the interaction of the cavity field, the mechanical mirror/membrane can be cooled down to temperatures corresponding to a state very close to the ground state[4]. Yet another but related system is nanomechanical resonators, which has been cooled down to the mK regime by means of coupling to single-electron transistors. At such low temperatures, quantum properties crucially begin to play an important role[5]. Both these mechanical systems pave the way to the study of quantum phenomena, such as entanglement and the superposition principle, on a macroscopic level[6].

Apart from exploring the borders between classical and quantum mechanics through entanglement of macroscopic objects[7], nanomechanical resonators provide also a way for multi-particle entanglement generation[8], a subject that has attracted a great deal of interest lately, see[9] and references therein. In particular, via piezoelectric interactions, nanomechanical oscillators can be coherently coupled to many Josephson junctions (JJs) simultaneously[10,11]. Indeed, an important problem associated with quantum computation and information is the problem of engineering entanglement in multi-particle systems. At the same time as controllable multi-particle entanglement schemes are necessary for quantum information processing, equally important is the understanding how decoherence affects these systems. Lately it has been shown that interesting aspects may arise when a system is coupled to an environment, different from the general idea that the reservoir will solely induce decoherence and thereby destroy entanglement. For example, Kim *et al.* demonstrated that initially uncorrelated quibits coupled to a thermal field can become entangled[12]. Later, Eberly and co-workers showed that the decay of entanglement shared between two initially entangled qubits, due to coupling to a reservoir, can become sudden, so called *entanglement sudden death*[13]. As entanglement might be lost instantaneously, it might also be regained suddenly, the *entanglement sudden birth*[14].

Motivated by, at the one hand the experimental progress in recent years, and at the other because of the need for a deeper understanding of the multi-particle entanglement, we study the dynamics of a nanomechanical oscillator coherently coupled to charge JJs. Losses of both the quibits as well as the resonator are taken into account by solving the full system-reservoir master equation. We find both entanglement sudden death and birth dynamics, i.e. environmentally generated entanglement. Furthermore, our master equation approach allows for the investigation of temperature dependence, and we especially find that the sudden birth of entanglement is relatively insensitive to the reservoir temperature. This is further concluded from the fact that the many subsystems (JJs and resonator) disentangle at specific times regardless of the initial state of the resonator. Due to the effective dipole-dipole coupling appearing between the JJs, the model has resemblances with spin $XX$ and $XY$ models[15] and thereby gives insight into decoherent evolution of generalized interacting spin systems. We, however, restrict our analysis, due to exponentially growth of the Hilbert space dimensions, to a maximum of three JJs coupled to the nanoresonator.

This paper is organized as follows. In Sec. II we derive the effective Hamiltonian of the system for arbitrary number of JJs. The general master equation is then introduced. Section III presents our results, where we in particular study concurrence and the tangle. The paper is concluded in Sec. IV.

(Color online) Schematic diagram of $N$ charge qubits coupled through the nanomechanical resonator. Here, each superconducting-quantum-interference-device loop (SQUID) connecting the superconducting island is represented by an effective JJ of coupling energy $E_J$ and capacitance $C_J$. The Cooper-pair box is biased by a voltage $V_g$ through the gate capacitance $C_g$ and coupled to the bulk superconductors by two identical small JJs and $\phi_c$ is the field-induced flux through the SQUID loop.

## II. MODEL SYSTEM

To begin with, we consider a system of $N$ charge qubits coupled to a nanoresonator via electrostatic interaction[16]. We denote by $\hat{a}^\dagger$ and $\hat{a}$ the creation and annihilation operators for a resonator phonon. Assuming weak or moderate vibration amplitudes $\hat{n} \equiv \hat{a}^\dagger \hat{a}$, and an energy splitting of the two lowest qubit states $|g\rangle$ and $|e\rangle$ to be resonant or quasiresonant with the single phonon energy, it is legitimate to restrict the analysis to a qubit sub-space only spanned by these two states[17]. Such a reduction of the Hilbert space brings out an effective dipole-dipole coupling between the qubits[17]. Turning to an interaction picture, the $N$-qubit Hamiltonian becomes[18]

$$\begin{aligned} \hat{H} &= \hbar \upsilon \hat{a}\hat{a}^\dagger + \sum_{j,i=1,i\neq j}^{N} \chi \hat{\sigma}_x^{(i)} \hat{\sigma}_x^{(j)} \\ &+ \sum_{j=1}^{N} \left\{ V \hat{\sigma}_z^{(j)} - \frac{1}{2} E_J \hat{\sigma}_x^{(j)} + \omega \left( \hat{a}^\dagger + \hat{a} \right) \hat{\sigma}_z^{(j)} \right\}. \end{aligned} \tag{1}$$

Here, $\chi$ is the effective dipole-dipole coupling strength, $V$ the gate voltage, $E_J$ is the coupling energy of the Josephson junction and $\omega = \frac{eC_J}{2(C_g+C_J)} \sqrt{\upsilon/2\hbar C_F}$, is the effective resonator-qubit coupling, where $e$ is the electron charge, $C_F$ is the capacitance parameter, which depends on the thickness of the junction, $C_J$ is the junction capacitance, $C_g$ is the gate capacitance which is externally controlled and used to induce offset charges on the island, $\upsilon$ is the resonator frequency (see Fig. 1). The $\sigma$-operators are the standard Pauli matrices in the qubit basis $|g\rangle$ and $|e\rangle$; $[\hat{\sigma}_\alpha, \hat{\sigma}_\beta] = 2i\varepsilon_{\alpha\beta\gamma}\hat{\sigma}_\gamma$, $\hat{\sigma}_z|g\rangle = -|g\rangle$ and $\hat{\sigma}_z|e\rangle = |e\rangle$. Note that we have further assumed that all parameters are independent on charge qubits, *e.g.* the qubits are taken identical and the gate voltage the same for each of them.

In the absence of the nanoresonator, the above Hamiltonian is a special case of a driven $XY$ spin model; the gate voltage has the meaning of an external field, $\chi$ the spin-spin coupling and $E_J$ a resonant driving of each spin[15]. Furthermore, once coupled to the resonator, the model is related to the Dicke model of quantum optics where it has been shown that the dipole-dipole interaction induces novel phases such as spin glasses[19].

To take resonator as well as qubit losses into account, we follow the standard approach by considering a Markovian master equation[20]. This gives an equation of motion for the full system density matrix $\rho$, and takes the form[21]

$$\begin{aligned} \frac{\partial \hat{\rho}}{\partial t} &= -\frac{i}{\hbar}[\hat{H}_{in}, \hat{\rho}] - \kappa \left( \hat{a}^\dagger \hat{a} \hat{\rho} - 2\hat{a}\hat{\rho}\hat{a}^\dagger + \hat{\rho}\hat{a}^\dagger \hat{a} \right) \\ &- \frac{1}{2} \sum_{i,j=1}^{N} \gamma_{ij} (1+\bar{N}) \left( \hat{\sigma}_+^{(i)} \hat{\sigma}_-^{(j)} \hat{\rho} + \hat{\rho} \hat{\sigma}_+^{(i)} \hat{\sigma}_-^{(j)} - 2\hat{\sigma}_-^{(i)} \hat{\rho} \hat{\sigma}_+^{(j)} \right) \end{aligned}$$

$$-\frac{1}{2}\sum_{i,j=1}^{N}\gamma_{ij}\bar{N}\left(\hat{\sigma}_-^{(i)}\hat{\sigma}_+^{(j)}\hat{\rho}+\hat{\rho}\hat{\sigma}_-^{(i)}\hat{\sigma}_+^{(j)}-2\hat{\sigma}_+^{(i)}\hat{\rho}\hat{\sigma}_-^{(j)}\right). \quad (2)$$

The first term gives the unitary time evolution, while the remaining ones correspond to the effective coupling to the reservoirs. We symbolize by $\overline{N}$, the number of thermal phonons at the resonator frequency $\nu$. Here we consider the case of a thermal bath without squeezing. In Eq. (2), $\kappa$ is the resonator decay rate, and the $\gamma_{jj}$'s are effective decay rates of qubit $j$, which will all be taken equal $\gamma_{jj} \equiv \gamma$. The qubits undergo as well a correlated decay with rates $\Gamma \equiv \gamma_{ij}(i \neq j)$, leading to further decoherence.

In the representation $\widetilde{\rho}(t) = \exp\left(\frac{i}{\hbar}Ht\right)\rho(t)\exp\left(-\frac{i}{\hbar}Ht\right)$, the equations of motion for the density matrix attain a simple form. For arbitrary initial states and using Eqs. (1) and (2), we numerically obtain the matrix elements of the density operator. For the special case of a system consisting of a single lossless qubit coupled to the resonator, one can obtain exact analytical forms of the diagonal matrix elements by imposing the *rotating wave approximation* in which rapidly oscillating terms are neglected.

## III. DYNAMICS

We proceed by obtaining the solutions of the master equation introduced in the previous Section. One of the important aspects, especially for open systems, is the development of quantum correlations. The rule of thumb is that such correlations are quickly diminish in the presence of decoherence, but we will demonstrate that this is not necessarily the case. To this end, in this Section we present a detailed study of entanglement formation for two qubits coupled to a nanoresonator.

For two-qubit systems, the concurrence is a measure of the entanglement for both pure and mixed states[23,24]. For a density matrix $\hat{\rho}(t)$, which represents the state of a bipartite system, the concurrence is defined as $C(\hat{\rho}) = \max[0, \lambda_1 - \lambda_2 - \lambda_3 - \lambda_4]$, where the $\lambda_i$'s are the non-negative eigenvalues, in decreasing order ($\lambda_1 \geq \lambda_2 \geq \lambda_3 \geq \lambda_4$), of the Hermitian matrix $\hat{\Upsilon} \equiv \sqrt{\sqrt{\hat{\rho}}\widetilde{\rho}\sqrt{\hat{\rho}}}$ and $\widetilde{\rho} = (\hat{\sigma}_y \otimes \hat{\sigma}_y)\,\hat{\rho}^*\,(\hat{\sigma}_y \otimes \hat{\sigma}_y)$. Here, $\hat{\rho}^*$ is the complex conjugate of the density matrix $\hat{\rho}$ when it is expressed in a fixed basis and $\hat{\sigma}_y$ represents the Pauli matrix in the same basis. The function $C(\hat{\rho})$ ranges from 0 for a separable state to 1 for a maximum entanglement. The tangle in a pure two qubit system is defined as the square of the concurrence[25,26]. Rungta *et al.*[27] defined the $I$-concurrence in terms of a *universal-inverter* which is a generalization to higher dimensions of the two spin flip operation. However, for arbitrary bipartite mixed states one should minimize the average squared pure state concurrence[28]. In this case, the generalized mixed state form of the tangle can be written as

$$\tau\left(\hat{\rho}\right) = 2 \min_{S_i} \sum_i \left(1 - Tr[(\hat{\rho}_a^{(i)})^2]\right), \tag{3}$$

where $\hat{\rho}_a^{(i)}$ is the marginal state for the $i^{th}$ term in the ensemble decomposition and $S_i$ is a convex combination of the pure states. It is interesting mentioning here that, when we consider the two-qubit case, Eq. (3) gives the same results as the concurrence.

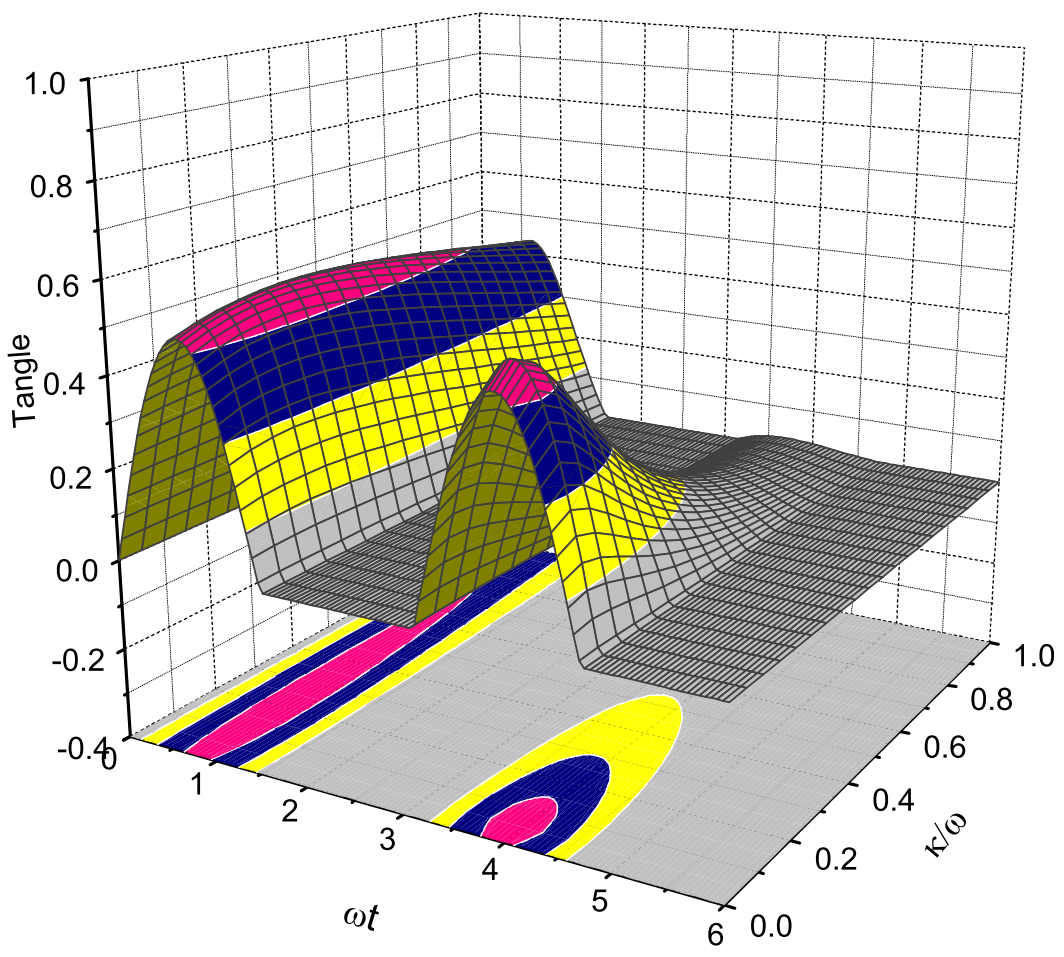

FIG. 1: (Color online) Time evolution of the tangle (3) between a single charge qubit and the resonator as a function of the scaled time $\omega t$ and resonator losses $\kappa/\omega$. The $JJ$ was initially in its excited state and the resonator in the vacuum state. The parameters are, $E_J = \nu$, $V/\omega = 1$, and $\gamma = \Gamma = 0.0$.

In the first scheme, we assume that a single charge qubit, i.e. $N = 1$, is coupled to the resonator which is damped at a rate $\kappa$, while the qubit losses, on the other hand, is assumed zero $\gamma = \Gamma = 0$ and $N_{th} = 0$. In this case, we consider the qubit initially in its excited state $|e\rangle$ and the initial state of the resonator is $|0\rangle$. The evolved global state $\hat{\rho}(t)$ is determined after finding the time evolution of each matrix element, by using Eq. (2). The resonator and qubit are two interacting systems, since the ranks of their reduced density operators $(\hat{\rho}_{r,q}(t))$ are not greater than two, $\hat{\rho}_{r,q}(t) = \mathrm{Tr}_{q,r}\left\{\widehat{\rho}(t)\right\}$. Hence, the use of tangle (concurrence) as a measure of quantum correlations/entanglement between the two subsystems applies in this case. In order to calculate the tangle, we choose the standard basis of the product space of each subsystem as $|0, e\rangle, |0, g\rangle, |1, e\rangle$ and $|1, g\rangle$ (here "1" indicates the one phonon Fock state), then one can effectively consider the density matrix of the system formed by two two-level subsystems at each instant of time $t$.

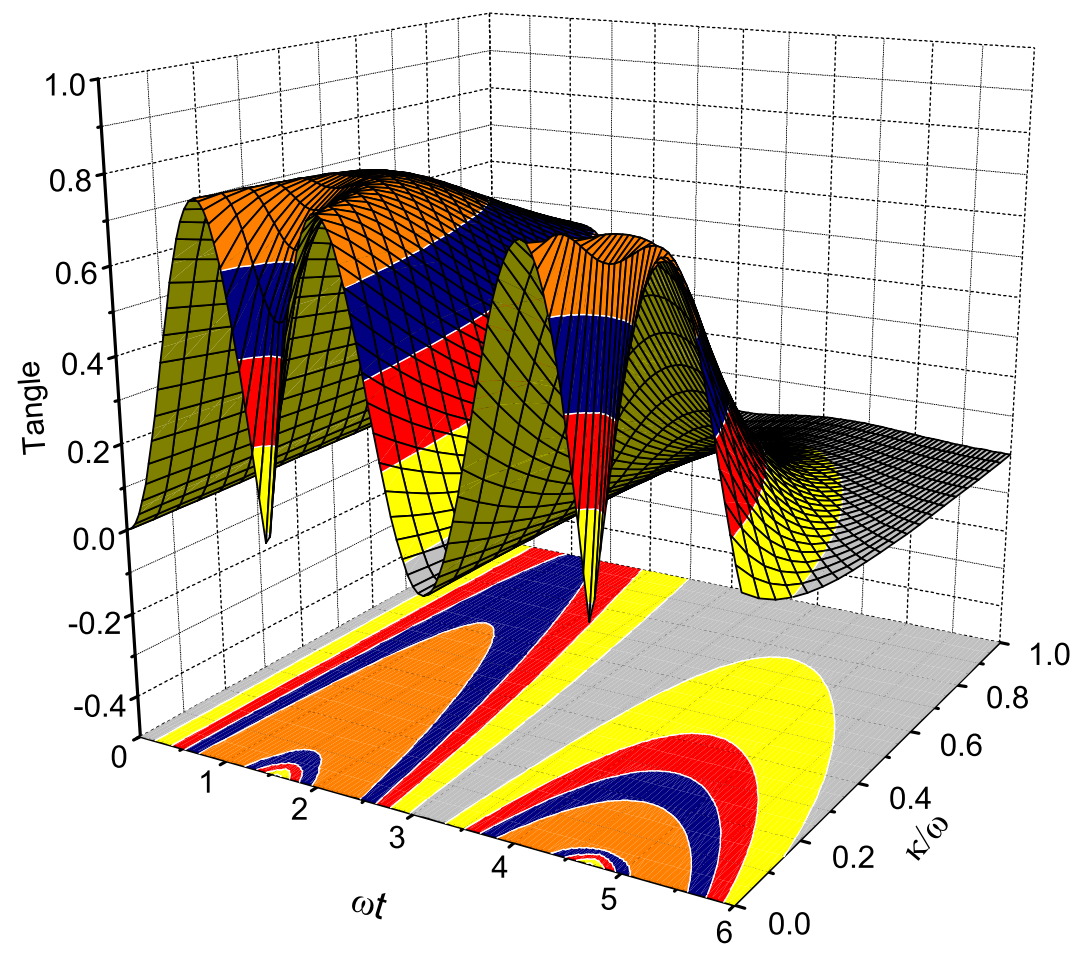

FIG. 2: (Color online) Time evolution of the tangle (3) between two charge qubits as a function of the scaled time $\omega t$ and resonator losses $\kappa/\omega$. The first qubit is initially in its excited state, the second qubit in its ground state, and the resonator in the vacuum state. The parameters are, $E_J = \nu$, $V/\omega = 1$, $\chi = 0$, and $\gamma = \Gamma = 0.0$.

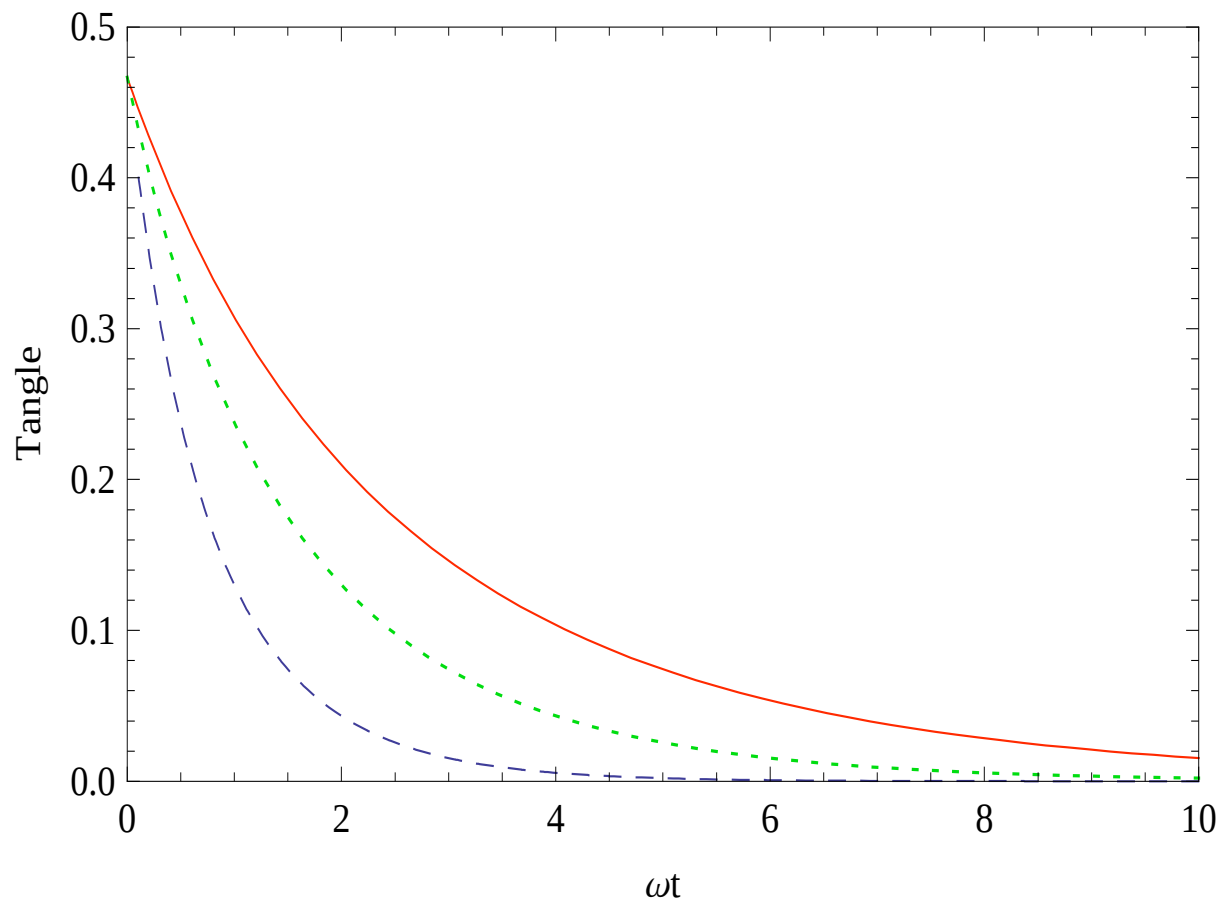

FIG. 3: (Color online) Tangle between two charge qubits as a function of the scaled time $\omega t$. The initial state of the resonator is a mixed state (see text for specific parameters) with different values of the decay rate $\gamma/\omega$, where $\gamma/\omega = 0.01$ (solid red curve), $\gamma/\omega = 0.1$ (dotted green curve) and $\gamma/\omega = 0.7$ (dashed blue curve). The other parameters are $\chi = 0$, $\kappa = 0$, $N_{th} = 0.5, \Gamma/\omega = 0.001$, and $E_J = \nu/\omega$.

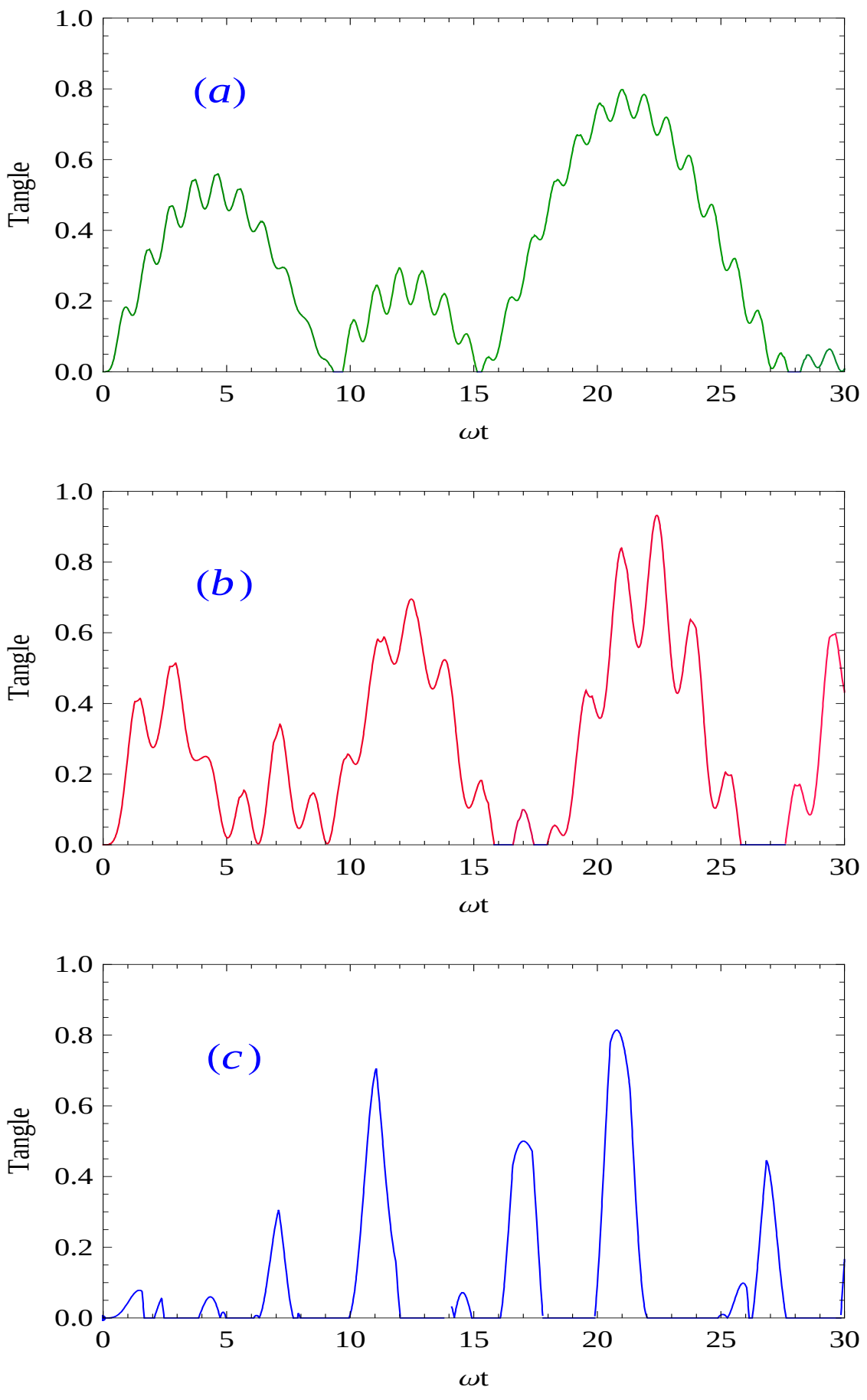

FIG. 4: Tangle between two charge qubits as a function of the scaled time $\omega t$. The initial state is $|\psi(0)\rangle = |e_1, e_2; 0\rangle$ with different values of $\chi/\omega$, where (a) $\chi/\omega = 30$, (b) $\chi/\omega = 15$ and (c) $\chi/\omega = 0.01$. The other parameters are the same as Fig. 3, but here, $\kappa = 0$ and $\gamma = \Gamma = 0.0$.

Figure I displays the tangle with dynamics especially showing a sequence of entanglement sudden deaths/births. Its evolution is strongly affected by the modulation term that depends on the decay rate $\kappa$. As the resonator decay rate increases, this effect is less pronounced and as expected, in the bad resonator limit all entanglement gets rapidly lost. It is seen that in the $\kappa \to 0$ limit we still have entanglement sudden deaths/births lasting over a large time span. Such features might seem remarkable since the von Neumann entropy (not presented here) in the rotating wave approximation is indeed a smooth function of time $t$ in the $\kappa = 0$ limit. It is known that even without the rotating wave approximation, the von Neumann entropy has a smooth time evolution[29]. To clarify this point, we stress the fact that, the von Neumann entropy measures entanglement of the reduced density matrix, while the tangle $\tau(\hat{\rho})$, used in this work, utilizes the evolved global state $\hat{\rho}(t)$ together with a reduced phonon basis in which the system substitutes two two-level subsystems. These aspects imply that the von Neumann entropy and the tangle measure different properties and thereby give different behaviors; one being smooth and one possessing discontinuous time derivatives.

In general, the time evolution is quite complicated. However, within the rotating wave approximation and the special case of $\chi = \gamma = \Gamma = 0$, taking into account the initial state of the system is $\hat{\rho}(0) = |e, 0\rangle\langle 0, e|$, the tangle takes a simple form,

$$\begin{aligned}\tau(\hat{\rho}) \;&=\; \max\left[0, \sin(2\omega t)\left\{2\cosh\left(\frac{\kappa}{\omega}\omega t\right) + \cosh\left(\frac{3\kappa}{\omega}\omega t\right)\right.\right. \\ &\left.\left. -2\sinh\left(\frac{\kappa}{\omega}\omega t\right) - \sinh\left(\frac{3\kappa}{\omega}\omega t\right)\right\}\right]. \qquad (4)\end{aligned}$$

In getting Eq. ( 4), we consider the temperature $T = 0$ K, i.e. the average thermal phonon number is zero. This expression verifies the fact that entanglement vanishes in the large decay limit $\kappa \to \infty$.

We now turn to the case of two qubits, opening up for multi partite entanglement. Thereby, since the qubits possess an indirect dipole-dipole interaction among each other, there is an additional term that drives the entanglement evolution. In the two qubit case, we assume the initial state of the qubits as

$$\begin{aligned}\rho^a(0) \;&=\; a|e,e\rangle\langle ee| + b|e,g\rangle\langle eg| + (1-a)|gg\rangle\langle gg| \\ &+f|e,g\rangle\langle ge| + f^*|ge\rangle\langle eg| + c|ge\rangle\langle ge|, \qquad (5)\end{aligned}$$

where $a, b, c$ and $f$ are constants. To compare with the single qubit case, we assume $a = 1$ and $b = c = f = 0$, i.e. both qubits start out in their excited states. From Fig. 2, we see that the problem of entanglement creation and evolution in the two-qubit system is not similar to the one-qubit case. In the present case, the entanglement does not show a zero value lasting for long times (compare Fig. 2 and Fig. 3). Also, for small values of the decay rate $\kappa$, additional drops in the tangle are seen. However, the entanglement vanishes in the large decay limit in a similar way as for the single qubit case.

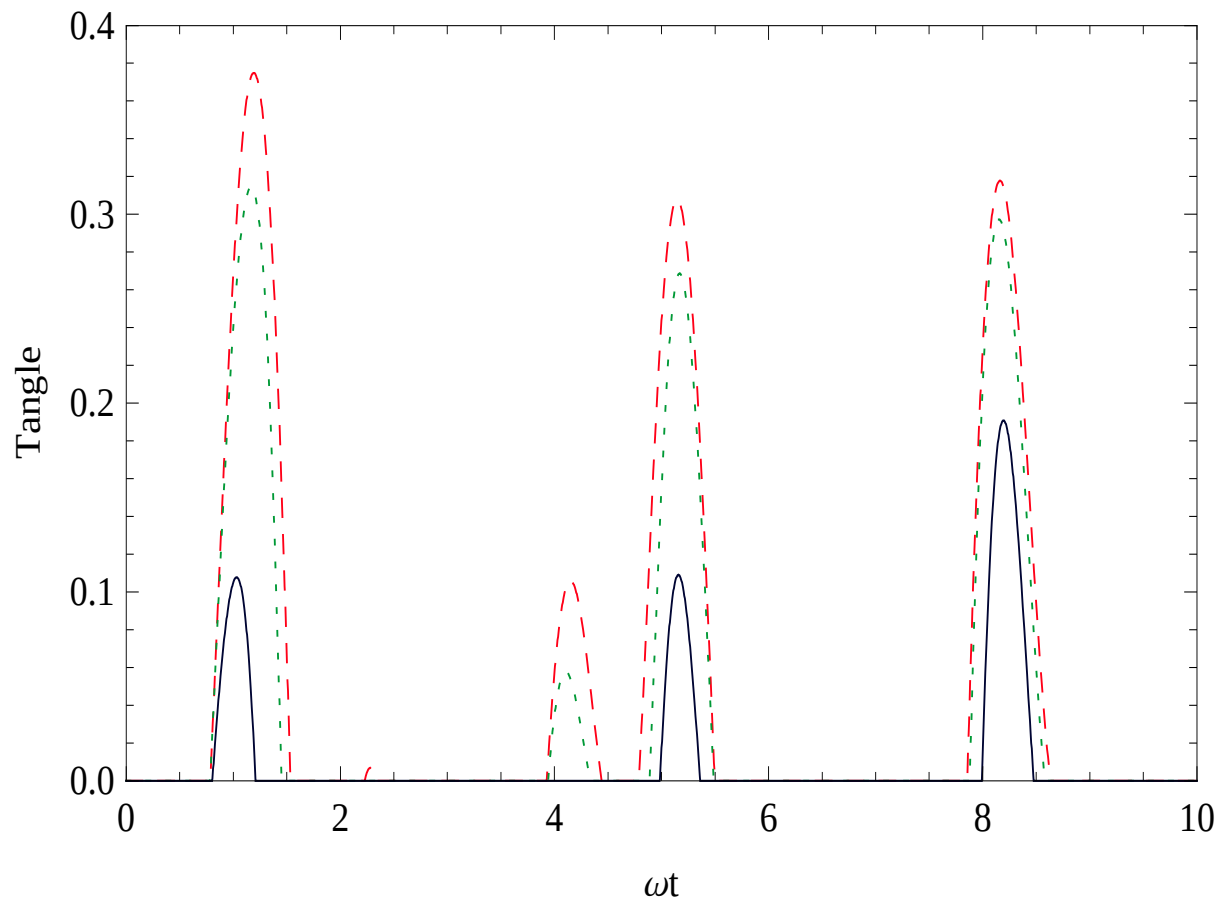

FIG. 5: (Color online) Tangle as a function of the scaled time $\omega t$ for the case with two initially excited charge qubits and different values of $\bar{N}$, where $\bar{N} = 0.01, 0.1$ and $0.5$, (dashed red, dotted green, and solid black lines respectively). The other parameters are the same as Fig. 3 and $\kappa = 0$.

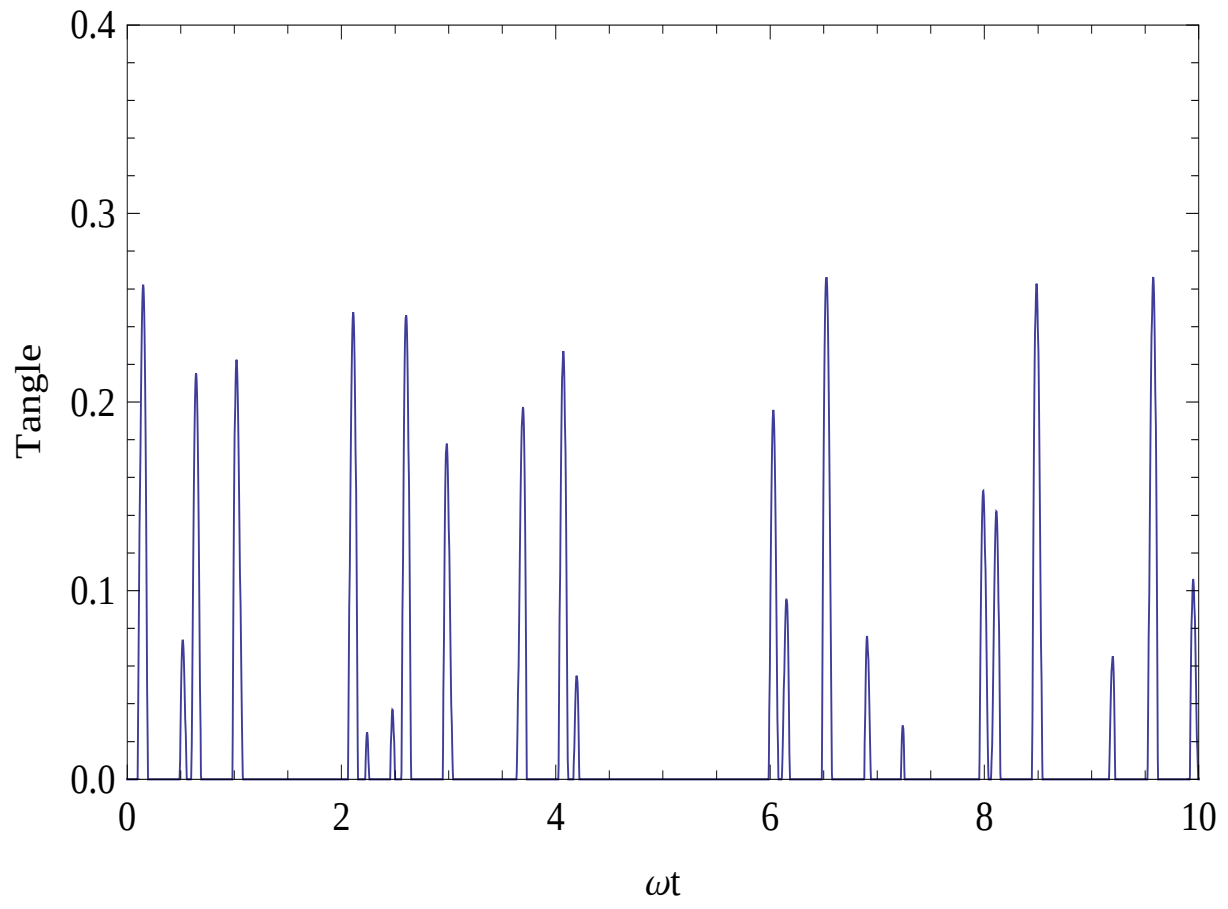

FIG. 6: Tangle as a function of the scaled time $\omega t$, (for the three qubit case). The initial state of the resonator is a thermal state with $\bar{N} = 0.5$,. The other parameters are the same as in Fig. 6. The initial qubit state is taken as $|eee\rangle$.

The time evolution of the tangle between two charge qubits for various values of the effective decay rate of the qubit $(\gamma/\omega)$ is displayed in Fig. 4. Here it is assumed that the initial state of the qubits is a mixed state, i.e. $a = 0.1$ and $b = c = 1 = 1$. In this figure, we see that the entanglement is sensitive for $\gamma/\omega$, but for sufficiently small values of $\gamma/\omega$ it is relatively robust and persists over longer time periods. However, for large values of $\gamma/\omega$, the sudden death of entanglement occurs.

In Fig. 5, the time evolution of the tangle between two qubits is plotted against the scaled time $\omega t$ for nonzero dipole-dipole coupling strengths ($\chi \neq 0$) when the qubits are initially in the state $|e_1, e_2\rangle$. In this figure we ignored the effect of losses $\kappa$ and $\gamma$. It is observed that the entanglement creation time decreases as $\chi$ becomes large. Which means that, this parameter can be used as a controller for entanglement creation, in agreement with You, et al[30] where they stated that this type of spin-spin coupling can be conveniently used to formulate an efficient quantum-computing scheme.

When the mean thermal phonon number is not zero, entanglement shows completely different time behavior. This is illustrated in Fig. 6, where we plot the tangle as a function of the scaled time for different values of the mean thermal phonon number. In contrast to the previous cases where entanglement was created immediately after $t = 0$, we note that the creation of entanglement is now delayed, that it occurs after a finite time. The reason for the delayed creation of entanglement can be understood as follows. When the system is prepared in the state $|e_1, e_2\rangle$, the resulting transitions are cascades: The system decays first to the intermediate states, from which it then decays to its ground state. Since the transition rates to and from the intermediate states are different when $\overline{N} \neq 0$, there appears unbalanced population distribution between these states. This may result in a transient entanglement between the qubits. Also, we see that in this case the entanglement sudden death always happens no matter which entangled state the qubits are initially in and no matter how small the nonzero mean thermal phonon number is. This implies that, even for relatively "hot" baths the system shows entanglement properties. Similar results were found in a system of two lossless two-level atom interacting with a finite $Q$ cavity[31].

Another important question one may ask: Does the properties of entanglement sudden birth depend on the number of qubits ($N > 2$), and especially is the effect less pronounced for more qubits? To answer this question, we plot in Fig. 7 the tangle with respect to the scaled time for three qubits ($N = 3$). This figure shows over a range of interaction times that the entanglement start to build up at earlier times closer to the initial interaction time. This is in striking contrast to the two qubits case displayed in Fig. 6 where the entanglement starts to build up at a much later time. Also, the number of sudden deaths/births is increased with less amplitudes. This might be interpreted as the state's entanglement becoming more robust when the size of the system increases. However, what really matters is not that the initial entanglement does not disappear fully but that a significant fraction of it remains, either to be directly used, or to be distilled without an excessively large overhead in resources. Overall, we find that the number of qubits acts as well as a control parameter for the qubit-qubit entanglement.

## IV. CONCLUSIONS

In this paper, we have investigated the time evolution of the entanglement between a nanomechanical resonator and $N = 1, 2, 3$ charge qubits. We have shown that by changing the dipole-dipole coupling, which can be achieved by varying the mutual distances between the JJs, one can control the entanglement. The decay which is regarded as a source of decoherence, creates in general entanglement sudden birth followed by a sudden death. Thus, entanglement may in fact be generated via the environment. Finally, the entanglement of three qubits is discussed. We have in particular demonstrated that by increasing the number of qubits, the entanglement starts at earlier times and shows more frequent sudden deaths/births.

The resilient properties of the system together with its entanglement robustness therefore makes it a suitable candidate for quantum information processing, fundamental studies of quantum many-body dynamics, or analyzes of the quantum-classical interface. Its nanomechanical structure might as well have applications in precision measurements[32].

**Acknowledgments**

We acknowledge useful collaborations and discussions with W. Ficek, D. Tsomokos and Arthur McGurn. JL acknowledges support from the MEC program (FIS2005-04627).